\documentclass[12pt]{iopart}

\usepackage{iopams}  
\usepackage{graphicx}

\newcommand{\ket}[1]{| #1 \rangle}
\newcommand{\bra}[1]{\langle #1 |}
\newcommand{\eq}[1]{Eq.~(\ref{#1})}
\newcommand{\fig}[1]{Fig.~\ref{#1}}

\begin{document}
\title{Boosting up quantum key distribution by learning statistics of practical single photon sources}

\author{Yoritoshi Adachi, Takashi Yamamoto, Masato Koashi and Nobuyuki Imoto}
\address{Division of Materials Physics, Department of Materials Engineering Science,
Graduate School of Engineering Science, Osaka University, 1-3 Machikaneyama, Toyonaka, Osaka 560-8531, Japan}
\ead{adachi@qi.mp.es.osaka-u.ac.jp}

\date{\today}

\begin{abstract}
We propose a simple quantum-key-distribution (QKD) scheme for
practical single photon sources (SPSs), which works even with a
moderate suppression
of the second-order correlation $g^{(2)}$ of the source. The scheme utilizes a passive
preparation of a decoy state by monitoring a fraction of the signal
via an additional beam
splitter and a detector at the sender's side to monitor photon number
splitting attacks.
We show that the achievable distance increases with the precision with which the
sub-Poissonian tendency is confirmed in higher photon number
distribution of the source,
rather than with actual suppression of the multi-photon emission events.
We present an example of the secure key generation rate in the case
of a poor SPS with
$g^{(2)} = 0.19$, in which no secure key is produced with the conventional
QKD scheme, and
show that learning the photon-number distribution up to several
numbers is sufficient for
achieving almost the same achievable distance as that of an ideal SPS.

\end{abstract}
\maketitle

A single-photon source (SPS), which emits exactly one photon in a well defined optical mode, has been actively studied with quantum dots \cite{KBKY99-Nature, Michler00-Science, Santori00-Nature, Yuan02-Science, BUARS05-OE, Kako06-NMat, ZAMZLFLGRZG06-APL, Strauf07-NPho, VZLA09-NPhys, FMRFDXMSS09-NPhys, AURLFM09-arXiv}, colour centres in diamond \cite{KMZW00-PRL, Gaebel04-NJP, Wu07-NJP, Simpson09-APL}, and other systems \cite{KLHLW04-Nature, McKeever04-Science, DJDBBSMBG05-Science, HWSWKR07-NPhys}. Behind those activities lies the fact that ideal SPSs serve as a useful resource for efficient quantum information processing such as quantum computation \cite{KLM01-Nature} and quantum key distribution (QKD) \cite{BB84-IEEE}.
But imperfections in practical SPSs affect the performance in those applications, especially in QKD \cite{Waks02-Nature, Beveratos02-PRL, Intallura07-APL} in which very rare emission events could be exploited by a potential eavesdropper Eve.
The imperfection that matters most in QKD is the
emission of multiple (two or more) photons in a single pulse, which
we assume to occur with probability $p_{\rm multi}$.
Usually the two-photon events are dominant in $p_{\rm multi}$, in
which case it is related to the normalised
second-order correlation $g^{(2)}$ by $p_{\rm multi}=\mu^2 g^{(2)}/2$
with $\mu$ being the average photon number emitted
in a pulse. Since $\mu\sim O(1)$ in practical SPSs, $p_{\rm multi}$
and $g^{(2)}$ are of the same order of magnitude.
In the BB84 QKD protocol, which has been proposed by Bennett and Brassard in 1984 \cite{BB84-IEEE}, whenever the sender Alice
emits multiple photons Eve can
steal a photon to obtain full information without introducing any
error. 
To make matters worse, we cannot exclude the
possibility that Eve may force the receiver Bob to detect a photon
preferentially in such a multi-photon emission event.
As a result of this photon-number splitting (PNS) attack, the protocol produces no secret key beyond
a threshold distance at which Bob's detection rate $Q$ is comparable to
$p_{\rm multi}$ \cite{BLMS00-PRL}. 
Unless we accept a significant reduction of the emission probability
$\mu$ \cite{WSY02-PRA}, it poses a severe requirement on $g^{(2)}$.
For example, for an optical channel with a loss of 0.2 dB/km of the typical telecommunication fibre and for
Bob's detection apparatus with efficiency 0.01, it is estimated under $\mu\sim O(1)$ that $g^{(2)}\sim O(10^{-6})$ is
needed to reach a communication distance of 200 km.
This requisite will be hard to achieve in real experiments, and if at
all, it may require sacrificing other performances
such as the repetition rate.

\begin{figure}[]
\begin{center}
\includegraphics[scale=1]{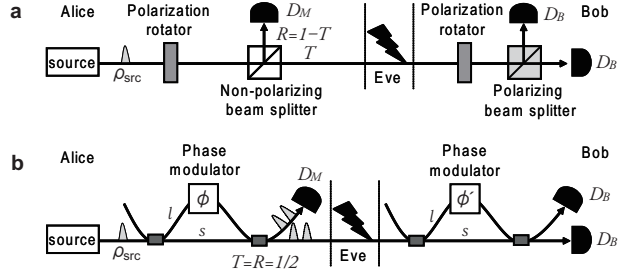}
\caption{
{\bf a,} Polarization-encoding BB84 QKD protocol with the proposed modification. The only difference from the conventional system is the addition of a non-polarizing beam splitter to measure a fraction $R=1-T$ of the signal with a monitoring detector $D_M$ at Alice's side.
{\bf b,} Phase-encoding BB84 QKD protocol based on a DMZI with the proposed modification. Alice uses two 50/50 fibre couplers and a phase modulator, and encodes bit information on the relative phase between two pulses. Bob uses the same setup as Alice's followed by two threshold detectors ($D_B$'s). The only difference from the conventional setup is just the use of a monitoring detector $D_M$ at Alice's side.
}
\label{fig:adachi1}
\end{center}
\end{figure}

In order to circumvent this problem, one may depart from the BB84 protocol to adopt a different sifting procedure \cite{SARG04-PRL} yielding a key even from the multi-photon emission events \cite{Koashi05-arXiv, TL06-PRA}. Another common direction is to keep the protocol itself and to devise a way to obtain a clue about photon number dependence of Eve's attack. Our main idea in this paper is to use the correlation between the two output lights from a beam splitter to obtain such a clue \cite{AYKI08-AQIS, CMML09-OL}.
For the BB84 protocol with polarization encoding, this idea is implemented by inserting a  non-polarizing beam splitter with a transmittance $T$ and placing a {\em monitoring} detector $D_M$ at Alice's side as in Fig.~\ref{fig:adachi1}~(a). For a standard phase-encoding system based on a double Mach-Zehnder interferometer (DMZI), mere addition of $D_M$ at the open port implements our scheme with $T=1/2$, as shown in Fig.~\ref{fig:adachi1}~(b).
If the light incident on $D_M$ is suitably correlated to the number of photons
sent toward Bob, we may detect Eve's PNS attacks like the decoy-state idea \cite{Hwang03-PRL, LMC05-PRL, Wang05-PRL} and particularly passive decoy-state generation schemes \cite{MS07-PRA, AYKI07-PRL, ML08-NJP}.

\begin{figure}[]
\begin{center}
\includegraphics[scale=1]{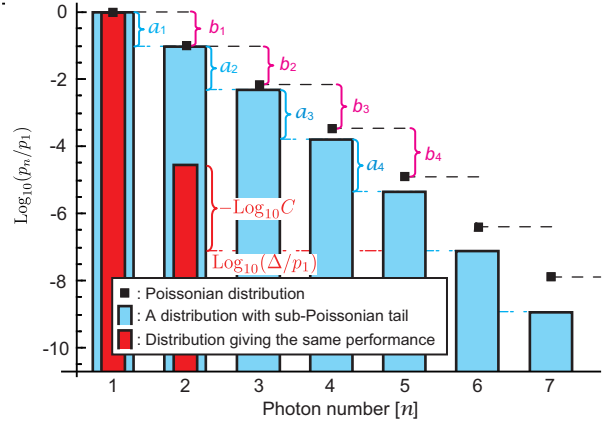}
\caption{
Our scheme works when the photon number distribution $\tilde{p}_n$
of the source has a ``sub-Poissonian tail,'' which has a simple meaning
in the usual cases with $\tilde{p}_1\gg \tilde{p}_2\gg \cdots$.
In the figure, the distribution is shown as blue bars in logarithmic scale,
together with square dots corresponding to a Poissonian distribution $\{P_n\}$
satisfying  $\tilde{p}_2/\tilde{p}_1=P_2/P_1$. The decrements are denoted by
$a_n\equiv {\rm log}_{10} (\tilde{p}_n/\tilde{p}_{n+1})$ and $b_n\equiv {\rm log}_{10} (P_n/P_{n+1})$. 
When the tail of the distribution vanishes more rapidly than the
Poissonian distribution, namely, when $a_n-b_n\ge r$ for all $n\ge 2$
with a positive constant $r>0$, our simple scheme can foil the PNS attacks
and the multi-photon emission becomes no concern ($p_{\rm multi}$ is effectively zero).
When $a_n-b_n\ge r$ is confirmed only up to $n=n_{\rm max}-1$ with the residual
uncharacterized portion $\Delta \cong \tilde{p}_{n_{\rm max}+1}$, the detection
of the PNS attacks becomes imperfect but our scheme still
achieves performance comparable to the use of a source
with $p_{\rm multi}=\Delta/C$, with $C$ being a constant proportional to
$1-10^{-r}$ (shown as red bars for $n_{\rm max}=5$).
}
\label{fig:adachi2}
\end{center}
\end{figure}

In order for the above strategy to work, knowledge of emission
statistics for higher photon numbers is essential. As we will show,
our scheme can detect the PNS attacks when the higher photon number
tail drops more steeply than a Poissonian distribution, as explained in \fig{fig:adachi2}. If this sub-Poissonian tendency continues to arbitrary large photon
numbers, Eve can no longer force Bob to detect a photon preferentially
in a multi-photon emission event. In practice, the characterization of the
emission statistics is not complete and will be left with an uncharacterized
portion $\Delta$. In the simplest case where Bob detects photons at rate $Q$ with no errors,
the key rate $G$ in our scheme is approximated as $G=Q-(\Delta/C)$, which should
be compared with the rate in the conventional scheme $G=Q-p_{\rm multi}$ \cite{BLMS00-PRL}.
  Here $C$ is a constant reflecting the degree of the sub-Poissonian tendency.
Since $Q$ scales as $10^{-\alpha l/10}$ with distance $l$, the threshold distance
at which $G=0$ in our scheme scales as $(-\log_{10} \Delta)-(-\log_{10} C) +const.$, whereas
it is $(-\log_{10} p_{\rm multi})+const.$ in the conventional scheme. Therefore,
without actually suppressing multi-photon emission $p_{\rm multi}$ by a factor, we obtain the
same improvement merely by reducing the ambiguity $\Delta$ in learning the statistics
by the same factor.

From now on, we focus on the case where $D_M$ is a 
threshold detector with dark count rate $d_M$ and efficiency $\eta_M$.
The outcome of this detector is binary: When $n$ photons are incident, 
it produces a click with probability $1-(1-d_M)(1-\eta_M)^n$, 
and otherwise gives no click. We classify the events to 
{\em clicked} events and {\em non-clicked} ones, according to the
outcome of $D_M$.

For the security analysis, it is convenient to describe the correlation 
between the two outputs of Alice's beam splitter through a set of parameters 
$\{\gamma_n\}$, defined to be the conditional probability of $D_M$
to produce a click, given that $n$ photons are sent toward Bob. 
With $(\eta_M,d_M,T)$ fixed, $\{\gamma_n\}$ depend solely on the 
density operator $\rho$ of the source, which we assume to take the 
form of $\rho=\sum_{n=0}^\infty p_n \ket{n}\bra{n}$.
Let us define a characteristic function $\kappa_\rho(t)$ of the source by
\begin{eqnarray}
 \kappa_\rho(t)\equiv {\rm Tr}[\rho t^{\hat{n}}]=\sum_{n=0}^\infty p_n t^n,
\end{eqnarray}
where $\hat{n}\equiv \sum_{n=0}^\infty n \ket{n}\bra{n}$. Then the conditional probability $\gamma_n(\rho)$ is conveniently represented using 
its $n$-th derivative $\kappa_\rho^{(n)}(t)\equiv d^n\kappa_{\rho}(t)/dt^n$ as
\begin{eqnarray}
 \gamma_n(\rho)=1-[(1-d_M)\kappa^{(n)}_\rho(R(1-\eta_M))/\kappa^{(n)}_\rho(R)],
 \label{gamma}
\end{eqnarray}
where $R=1-T$.
When $\{p_n\}$ is Poissonian with mean $\mu$, it follows that $\kappa_\rho(t)=e^{-\mu(1-t)}$ and hence $\gamma_n(\rho)$ is independent of $n$. For $d_M\ll 1$ and $p_1\gg p_2\gg p_3\gg \cdots$, the parameters $\gamma_n(\rho) (n\ge 1)$ are approximated by
\begin{eqnarray}
 \gamma_n(\rho) \sim R\eta_M (n+1)p_{n+1}/p_n.
 \label{approximation}
\end{eqnarray}
Hence, roughly speaking, the values of $\{\gamma_n(\rho)\}$ represent how slowly the tail of the photon number distribution vanishes. For example, if $\gamma_1(\rho)>\gamma_n(\rho)$
holds for all $n\ge 2$, the tail of the distribution vanishes more rapidly compared to the Poissonian distribution with the same ratio $p_2/p_1$. In this paper, we say such a distribution has a ``sub-Poissonian tail.''

In fact, a source with a sub-Poissonian tail is ideal for our present scheme to work. Let 
\begin{eqnarray}
Q=Q^{[c]}+Q^{[nc]}
\label{relation}
\end{eqnarray}
be the overall rate of detection by Bob, where the superscript $c$ stands for the clicked events, and $nc$ for the non-clicked events.
These quantities are actually observed in the protocol. On the other
hand, 
$Q$ is also decomposed as $Q=\sum_n Q_n$, where $Q_n$ stands for the portion in which Alice has sent out $n$ photons toward Bob (we call it ``$n$-photon'' events henceforth). Since $D_M$ clicks with probability $\gamma_n(\rho)$ in the 
$n$-photon events,
we should have $Q^{[c]}=\sum_n \gamma_n(\rho) Q_n$ 
in the asymptotic limit of a large number of pulses.
Then, if Eve substitutes $1$-photon events by $n$-photon events with 
$n\ge 2$, $Q^{[c]}$ decreases due to the condition 
$\gamma_1(\rho)>\gamma_n(\rho)$ $(n \geq 2)$ and the attack will be detected.
She may try to increase $0$-photon events to compensate 
the decrease in $Q^{[c]}$, but it will then increase the observed
error rate, since if Alice emits no photon, the probability of bit
errors is $1/2$. Therefore, a source strictly satisfying 
 $\gamma_1(\rho)>\gamma_n(\rho)$ $(n\ge 2)$ allows us to detect PNS
 attacks in a very simple way.

In practice, it may be hard to find such a source, and it is even harder 
to prove it by characterizing 
the photon-number distribution of the source. For this reason, 
here we relax the condition and consider an approximate version of 
a source with a sub-Poissonian tail as follows.

{\it Assumption on the source} --- The density operator $\rho_{\rm src}$
 of a pulse from the source is written as a mixture of two
normalised density operators as
\begin{eqnarray}
 \rho_{\rm src}=(1-\Delta)\rho_{\rm sp}+\Delta\rho_{\rm uk},
 \label{source}
\end{eqnarray}
where $0\le \Delta < 1$. With an integer $n_{\rm max}$,
which may be infinity, $\rho_{\rm sp}$ is written in the 
form  $\rho_{\rm sp}=\sum_{n=0}^{n_{\rm max}} p_n \ket{n}\bra{n}$.
There exists $\Gamma$ satisfying 
\begin{eqnarray}
 \gamma_1(\rho_{\rm sp})>\Gamma\ge \gamma_n(\rho_{\rm sp})
 \label{condition}
\end{eqnarray}
for $2\le n \le n_{\rm max}$.

Under this assumption, the source emits an arbitrary unknown pulse 
with a (small) probability $\Delta$, but otherwise the distribution 
has a sub-Poissonian tail, whose degree is measured by parameter 
$\Gamma$. We expect that a source will satisfy this assumption in 
the following scenario, for example. A pseudo-single-photon source
with distribution $\{\tilde{p}_n\}$ is given, and a sequence of 
characterization experiments tells us good estimates of probabilities 
$\tilde{p}_0, \tilde{p}_1, \ldots, \tilde{p}_{n_{\rm max}}$. Then we choose
$\Delta=\sum_{n=n_{\rm max}+1}^\infty \tilde{p}_n$ and 
$p_n= \tilde{p}_n/(1-\Delta)$ for $n\le n_{\rm max}$.
After calculating the values of $\{\gamma_n(\rho_{\rm sp})\}$ 
from $p_0, p_1, \ldots, p_{n_{\rm max}}$,
we take 
$\Gamma=\max\{\gamma_n(\rho_{\rm sp})|2\le n \le n_{\rm max}\}$
and see if $\gamma_1(\rho_{\rm sp})>\Gamma$ holds.
In this example, $\rho_{\rm uk}$ stands for the uncharacterized portion
of the source. We may also use $\rho_{\rm uk}$ to absorb the higher
photon-number part responsible for an unwanted relation 
$\gamma_1(\rho_{\rm sp})<\gamma_n (\rho_{\rm sp})$.

For a source with $\Delta > 0$, Eve may be able to increase 
$Q^{[c]}$ by letting Bob preferably detect 
the events from $\rho_{\rm uk}$. But the amount of the increase is 
obviously no larger than $\Delta$. This increase leaves Eve a room
for substituting 1-photon events with multi-photon events with 
rate up to $\Delta/(\gamma_1(\rho_{\rm sp})-\Gamma)$, but no more.
Hence, as long as the overall detection rate is higher than this rate,
one may still expect to generate a secret key. 
In Appendix, we confirm this in a rigorous analysis.

Before we give a numerical example showing the improvement in
the secure key rate, we discuss the ideal case where Bob has observed no
bit errors. In this case, the secure key rate is given by 
$[Q^{[c]}-\Gamma Q -\Delta(1-\Gamma)]/(\gamma_1({\rho}_{\rm sp})-\Gamma)$ (see Appendix).
Since Alice uses a pseudo-single-photon source, the single-photon contribution should be dominant in Bob's detection events, namely, $Q^{[c]}/Q \simeq \gamma_1({\rho}_{\rm sp})$. Using the fact $\Gamma \ll 1$, the key rate is then simplified as $Q-\Delta /(\gamma_1({\rho}_{\rm sp})-\Gamma)$, 
confirming the dependence which we promised in the
introductory part with $C=\gamma_1({\rho}_{\rm sp})-\Gamma$.
This means that the key rate is positive as long as 
$Q > \Delta /(\gamma_1({\rho}_{\rm sp})-\Gamma)$, implying that the
uncharacterized portion $\Delta$ mainly determines the threshold
distance at which the key rate vanishes.  

\begin{figure}[]
\begin{center}
\includegraphics[scale=1]{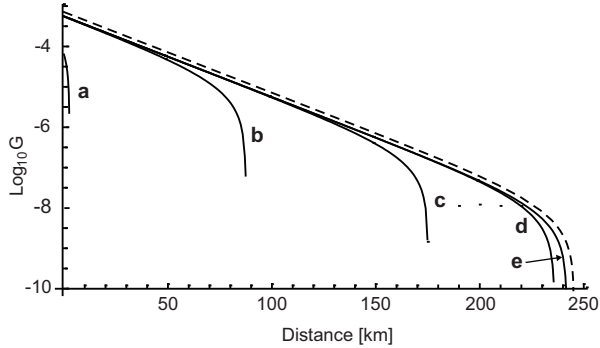}
\caption{
Alice's source $\rho_{\rm src}$ is chosen to be a mixture of a single-photon source with a loss and Poissonian background. The distribution is given by  $\tilde{p}_0 = p_0^{\rm (sp)} P_0^{\rm (Poi)}$ and $\tilde{p}_n = p_0^{\rm (sp)} P_n^{\rm (Poi)} + p_1^{\rm (sp)} P_{n-1}^{\rm (Poi)}$ for $n\geq 1$, with $P_n^{\rm (Poi)}\equiv e^{-\nu}\nu^n/n!$. We set $p_0^{\rm (sp)}=0.1$, $p_1^{\rm (sp)}=0.9$, and $\nu=0.1$, which leads to $p_{\rm multi}=0.086$ and $g^{(2)}=0.19$. (Its distribution is plotted as the blue bars in \fig{fig:adachi2}.)
The residual parameters are chosen as follows: $\eta_M=0.15$; $d_M=10^{-6}$; the error correction efficiency, $f=1.2$; loss coefficient of channel, $0.2$ [dB/km]; error rate of channel, $0.03$ (independent of distance). Each of Bob's detectors has the same dark count rate $2.5\times 10^{-9}$ and efficiency $0.01$ \cite{Takesue07-NPho}. {\bf a,}  $n_{\rm max}=4$, {\bf b,} $n_{\rm max}=5$, {\bf c,} $n_{\rm max}=6$, {\bf d,} $n_{\rm max}=7$, {\bf e,} $\Delta=0$. 
We see that increasing $n_{\rm max}$ leads to a constant improvement of the
threshold distance, until the overall detection rate drops to the order of Bob's dark count rate. Dashed curve shows the rate in a standard phase-encoding system (without $D_M$) based on a DMZI with an ideal SPS.
}
\label{fig:adachi3}
\end{center}
\end{figure}

We further show that the tendency in the ideal cases discussed above is also qualitatively valid with realistic channels and detectors, for an example of pseudo-single-photon source which is modelled as an ideal SPS with a loss and Poissonian background. In \fig{fig:adachi3}, we have chosen a source with a rather high multi-photon emission rate of 0.086 ($g^{(2)}=0.19$), which cannot generate secret key at any distance in the conventional scheme. Figure~\ref{fig:adachi3} shows the key rates for the same source with varied levels of characterization. For this source with $p_{n+1}/p_n \sim 10^{-1.7}$, increasing the maximum photon number $n_{\rm max}$ by one reduces $\Delta$ by factor of $10^{1.7}$. The analysis for the ideal case above suggests that this will increase the threshold distance by $17$ dB/(0.2 dB/km) = 85 km, which agrees with the curves in Fig.~\ref{fig:adachi3} for shorter distances where the error rate is not so high.

Throughout this paper, we only considered the limit of a large number of repetitions and a large final key, in which errors in estimating parameters due to statistical fluctuations are negligible. In practice, the key has a finite size and we must take the statistical fluctuations into account. This is beyond the scope of this paper but a few remarks are in order. Although our scheme may appear to involve a lot of parameters, it requires only a few of them to be estimated while actually running the protocol, due to its simplicity. The rest of the parameters, namely, the photon-number statistics of Alice's source, can be characterized off-line, before running the actual protocol. Of course, a care must be taken in this characterization to allow for practical issues such as the stability of the source.

Finally, we discuss the opposite condition of \eq{condition}, $\gamma_1(\rho_{\rm sp}) < \Upsilon \leq \gamma_n(\rho_{\rm sp})$ for $2\leq n \leq n_{\rm max}$.
This means a ``super-Poissonian tail," namely, a tendency that if multiple photons appear at one of the two outputs of the beam splitter, then the other one has a larger probability of having photons. For example, the heralded parametric down-conversion (PDC) source shows this tendency, which is ascribed to the thermal photon-number distribution observed in the signal mode of PDC. In this case, $\gamma_1(\rho_{\rm sp})$ is the minimum of $\{\gamma_n(\rho_{\rm sp})\}$ and any attempt by Eve to block 1-photon events should show up as an increase in the observed value of $Q^{[c]}$. Therefore we can also obtain secure key by using a similar analysis. We found that the rule of thumb for the key rate in this case is again
given by $G=Q-\Delta/C$, with $C=1-\gamma_1(\rho_{\rm sp})/\Upsilon$.


In conclusion, we have proposed a simple scheme to detect eavesdropping
attacks on multi-photon emissions from practical single-photon sources.
In contrast to the conventional scheme in which actual suppression of
multi-photon emissions is necessary to avoid such attacks, our scheme
requires no suppression as long as the higher photon-number tail of
emission probabilities decreases more rapidly than a Poissonian distribution.
As an example, we calculated the key rate for a lossy single-photon source
suffering from Poissonian background noise, which is regarded as a
bad source in terms of $g^{(2)} $. We found that reducing the
ambiguity in the characterization of the source improves the key rate,
in much the same way as the actual suppression of $g^{(2)} $ does in the
conventional scheme. 
It is worth mentioning here that 
the requisite in our scheme is satisfied by SPSs following a wide range of simple theoretical models: Any
source with imperfection modelled by loss and Poissonian background satisfies the requisite of 
sub-Poissonian tail, regardless of the amount of the loss and the background. On the other hand, if the background noise follows a thermal distribution, the source always has a super-Poissonian tail, which also meets our alternative condition discussed above. This also opens up a possibility that we may intentionally increase the fluctuations in the background to meet the requisite. 
Naturally, our result provides a new perspective on the desired performance of a practical single-photon source, namely, property of photon number correlations of orders higher than $g^{(2)} $. 
Recent development in the streak cameras \cite{WGJABBKRSHFKKH09-Nature} and photon-number-resolving detectors \cite{KTYH99-APL, RLMN05-PRA, KYS08-NPho} are helpful in studying such higher-order correlations. Therefore 
it will be interesting to experimentally characterize them for the single-photon sources currently under development, as well as to seek plausible theoretical models of high photon-number emissions.
\\

\section*{Acknowledgments}

We thank Hoi-Kwong Lo for helpful discussions. This work was supported by JSPS  Grant-in-Aid for Scientific Research(C) 20540389 and by MEXT Grant-in-Aid for Scientific Research on Innovative Areas 20104003 and 21102008, Global COE Program and Young scientists(B) 20740232. YA is supported by JSPS Research Fellowships for Young Scientists.

\appendix
\setcounter{section}{1}
\section*{Appendix}

Since one can simulate the source $\rho_{\rm src}$ equivalently 
by actively switching between the sources $\rho_{\rm sp}$
and $\rho_{\rm uk}$, we are allowed to label each detection event 
according to which of the source was used. As a result, 
the overall detection rate $Q$ is decomposed as 
$Q=\chi+Q_{\rm uk}$, where $\chi$ 
(used instead of $Q_{\rm sp}$ for simplifying the notations)
is the fraction with 
the source $\rho_{\rm sp}$, and $Q_{\rm uk}$ is with $\rho_{\rm uk}$.
Using similar decomposition for $Q_n$, we have
\begin{eqnarray}
Q=\chi_0+\chi_1+\chi_{\ge 2} + Q_{\rm uk},
\label{eq:overall_decomp}
\end{eqnarray}
where $\chi_{\ge 2}\equiv \sum_{n=2}^{n_{\rm max}} \chi_n$. 
The clicked and non-clicked portions are also written as 
$(j=c, nc)$
\begin{eqnarray}
Q^{[j]}=\chi_0^{[j]}+\chi^{[j]}_1+\chi^{[j]}_{\ge 2} + 
Q^{[j]}_{\rm uk}.
\end{eqnarray}
Since $\chi_n^{[c]}=\gamma_n(\rho_{\rm sp})\chi_n$ and 
$Q^{[c]}_{\rm uk}\le Q_{\rm uk}$, we use \eq{condition} to have 
\begin{eqnarray}
Q^{[c]}\le \gamma_0(\rho_{\rm sp})\chi_0+
\gamma_1(\rho_{\rm sp})\chi_1+\Gamma \chi_{\ge 2}+
Q_{\rm uk}.
\end{eqnarray}
Eliminating $\chi_{\ge 2}$ using Eq.~(\ref{eq:overall_decomp}) and
noting that $Q_{\rm uk}\le \Delta$, we obtain a bound on $\chi_1$
as
\begin{eqnarray}
&\chi_1 & \ge \xi(\chi_0)
\equiv (\gamma_1({\rho}_{\rm sp})-\Gamma)^{-1}
\nonumber \\
&&\times
[Q^{[c]}-\Gamma Q -(1-\Gamma)\Delta 
-(\gamma_0({\rho}_{\rm sp})-\Gamma)\chi_0].
\label{eq:xi}
\end{eqnarray}

Next, we derive an upper bound on the error rate $e_1$ 
for the 1-photon events. Let $E^{[c]}$ and $E^{[nc]}$ be
the overall error rates for the 
clicked events and the non-clicked events, respectively,
which can be directly estimated in the protocol.
Considering the contribution from 
 $0$- and $1$-photon events, we have, for $j=c, nc$,
\begin{eqnarray}
 Q^{[j]}E^{[j]}\ge \chi^{[j]}_0/2 + \chi^{[j]}_1 e_1.
\end{eqnarray}
Combining with 
$\chi_n^{[c]}=\chi_n-\chi_n^{[nc]}$,
we obtain $e_1 \le \epsilon(\chi_0)$ with
\begin{eqnarray}
&& \epsilon(\chi_0) \equiv \min \big\{
(Q^{[c]}E^{[c]}-\gamma_0\chi_0/2)/
[\gamma_1\xi(\chi_0)],
\nonumber \\
&& 
[Q^{[nc]}E^{[nc]}-(1-\gamma_0)\chi_0/2]/
[(1-\gamma_1)\xi(\chi_0)]
\big\},
\end{eqnarray}
where $\gamma_n=\gamma_n(\rho_{\rm sp})$. The allowed range of the 
parameter $\chi_0$ is determined from the condition
$\epsilon(\chi_0)\ge 0$.

The secure key generation rate for the clicked events or the non-clicked 
 events is given by the so-called GLLP formula \cite{GLLP04-QIC, Lo05-QIC, Koashi06-arXiv} as
\begin{eqnarray}
G^{[j]}/q = - Q^{[j]} f H_2(E^{[j]})+ Q_0^{[j]} 
&+Q_1^{[j]}[1-H_2(e_1)]
\label{eq:GLLP}
\end{eqnarray} 
$(j=c,nc)$, where $q$ denotes the protocol efficiency (e.g. $q = 1/2$ in
the standard BB84 protocol because half of the events are discarded 
by basis mismatch),
$f \geq 1$ the error correction efficiency, and $H_2(x) \equiv -{\rm
log}_2x-(1-x){\rm log}_2(1-x)$ the binary entropy function. 
Noting that $Q_n^{[j]}\ge \chi_n^{[j]}$ and assuming the worst case for
the choice of unknown parameter $\chi_0$, we have an expression for
achievable key rates as 
\begin{eqnarray}
G^{[c]}/q &=&  - Q^{[c]} f H_2(E^{[c]}) 
\nonumber \\
&& +\min_{\chi_0} \big \{ \gamma_0\chi_0 + 
\gamma_1\xi(\chi_0)[1-H_2(\epsilon(\chi_0))] \big\},
\nonumber \\
G^{[nc]}/q &=&  - Q^{[nc]} f H_2(E^{[nc]}) 
+\min_{\chi_0} \big \{ (1-\gamma_0)\chi_0  
\nonumber \\
&& + (1-\gamma_1)\xi(\chi_0)[1-H_2(\epsilon(\chi_0))] \big\},
\label{eq:R-each}
\end{eqnarray} 
where $\gamma_n=\gamma_n(\rho_{\rm sp})$.

When the secure key can be obtained from both events,
we can achieve a key rate better than $G^{[c]}+G^{[nc]}$ by 
 doing the error correction separately but the privacy amplification
 jointly.
 The achievable rate in this case is given by 
\begin{eqnarray}
G^{[both]}/q& = & - Q^{[c]}f H_2(E^{[c]}) -Q^{[nc]}f H_2(E^{[nc]})
\nonumber \\
&& \displaystyle + \min_{\chi_0} \{\chi_0 + \xi (\chi_0) [1-H_2(\epsilon (\chi_0))]\}.
\label{eq:R-both}
\end{eqnarray} 
Thus, the final key rate is given by $G = \max \{ G^{[both]}, G^{[c]}, G^{[nc]}\}$. In the ideal case where Bob has no bit errors ($E^{[c]}=E^{[nc]}=0$), we have $\chi_0=0$ and the final key rate is given by $G/q = \xi(0)=[Q^{[c]}-\Gamma Q -\Delta(1-\Gamma)]/(\gamma_1({\rho}_{\rm sp})-\Gamma)$.


\end{document}